\documentclass[journal]{IEEEtran}

\usepackage{lineno,hyperref}
\usepackage{bm}
\usepackage{mathrsfs}
\usepackage{subfigure} 
\usepackage{amssymb}
\usepackage{amsmath}
\usepackage{graphicx}
\usepackage{multirow}
\usepackage{textcomp,booktabs}
\usepackage[usenames,dvipsnames]{color}
\usepackage{colortbl}
\definecolor{mygray}{gray}{.9}
\usepackage{epstopdf}
\usepackage{cite}
\modulolinenumbers[5]

\ifCLASSINFOpdf

\else

\fi

\begin{document}
	
	\title{Robust Andrew's Sine Estimate Adaptive Filtering}
	
	\author{Lu Lu,~\IEEEmembership{Member,~IEEE}, Yi Yu,~\IEEEmembership{Member,~IEEE}, Zongsheng Zheng,~\IEEEmembership{Member,~IEEE},\\Guangya Zhu,~\IEEEmembership{Member,~IEEE}, and Xiaomin Yang,~\IEEEmembership{Member,~IEEE}
		\thanks{Manuscript received January 31, 2022. The work is supported by the National Science Foundation of P.R. China  under Grant 61901285 and 61901400, the
	Science Foundation of Sichuan Science and Technology Department under

			Grant 2021YFH0119 and 2023NSFSC0451, and the Fundamental Research Funds for the Central Universities under Grant 2021SCU12063.}
		\thanks{L. Lu and X. Yang are with the College of Electronics and Information Engineering, Sichuan University, Chengdu 610065, China. (e-mail: lulu19900303@126.com, arielyang@scu.edu.cn).} 
		\thanks{Y. Yu is with the School of Information Engineering, Southwest University of Science and Technology, Mianyang 621010, China. (e-mail: yuyi\_xyuan@163.com).} 
		\thanks{Z. Zheng is with the College of Electrical Engineering, Sichuan University, Chengdu 610065, China. (e-mail: zongsheng56@126.com).}
		\thanks{G. Zhu is with the High Voltage Laboratory in College of
Electrical Engineering, Sichuan University, Chengdu 610065, China. (e-mail:
miyazhu\_1989@126.com).}  
}
	
\markboth{}
{Shell \MakeLowercase{\textit{et al.}}: Bare Demo of IEEEtran.cls for IEEE Journals}

\maketitle

\begin{abstract}
The Andrew's sine function is a robust estimator, which has been used in outlier rejection and robust statistics. However, the performance of such estimator does not receive attention in the field of adaptive filtering techniques. Two Andrew's sine estimator (ASE)-based robust adaptive filtering algorithms are proposed in this brief. Specifically, to achieve improved performance and reduced computational complexity, the iterative Wiener filter (IWF) is an attractive choice. A novel IWF based on ASE (IWF-ASE) is proposed for impulsive noises. To further reduce the computational complexity, the leading dichotomous coordinate descent (DCD) algorithm is combined with the ASE, developing DCD-ASE algorithm. Simulations on system identification demonstrate that the proposed algorithms can achieve smaller misalignment as compared to the conventional IWF, recursive maximum correntropy criterion (RMCC), and DCD-RMCC algorithms in impulsive noise. Furthermore, the proposed algorithms exhibit improved performance in partial discharge (PD) denoising.
\end{abstract}


\section{Introduction}
\label{sec:intro}

\IEEEPARstart{T}{he} recursive least-squares (RLS) algorithm is challenged by the presence of outliers and high computational complexity \cite{poulo2008adaptive,farahmand2012robust,wang2011low}. Moreover, the standard RLS algorithm still has the instability problem because of the inverse operation \cite{arablouei2011modified}. In \cite{bhotto2012new,lu2018recursive,paleologu2008robust,slavakis2019robust,qian2019recursive}, robust RLS algorithms have been considered as insensitivity to noise, and in \cite{zhang2016convex} the RLS algorithm is incorporated into the maximum correntropy criterion (MCC), resulting in the recursive MCC (RMCC) algorithm to achieve enhanced performance. 

Recently, the iterative Wiener filter (IWF) \cite{xi2013iterative} algorithm was developed as an effective alternative to the RLS algorithm. Such algorithm is derived from the Wiener-Hopf equation and utilizes the variable step size (VSS) scheme for performance improvement. It speeds up the convergence rate of the least mean square (LMS) algorithm and significantly reduces the computational load of the RLS algorithm. To exploit the sparsity in channels, \cite{lim2020} provides the $L_1$-norm sparse penalty method based on the IWF algorithm. Based on this algorithm, \cite{lim2020maximum} employs the MCC to limit the effect of outliers, thus ensuring robustness against impulsive noise. However, the performance of the existing IWF algorithms still has a lot of room for improvement. Additionally, the computational complexity of the IWF-type algorithms can be further reduced.  

On the other hand, the leading dichotomous coordinate descent (DCD) method is an efficient improvement for solving the auxiliary normal equation as well as reducing the computational complexity of the RLS algorithm \cite{zakharov2008low}. By setting the quantization step size to the power of two, all divisions and multiplications can be transformed into addition and shift operations, thereby saving considerable cost \cite{chen2020efficient}. During the past decade, several
low-cost algorithms were developed based on DCD method, e.g., see \cite{chen2020efficient,yu2019dcd,zakharov2013dcd,nascimento2016rls,zakharov2008low,elisei2019recursive,zakharov2017low} and the references therein.

In this paper, we propose two novel recursive algorithms that use the Andrew's sine estimator (ASE). Before that, the ASE has been investigated in robust statistics, which reveals its effectiveness to recover outlier process \cite{andrews1974robust}. However, there is no literature focused on using the ASE for adaptive filtering. For low-power noise or outliers, the ASE performs similar to the $L_2$-norm estimator, while it behaves similar to the $L_1$-norm estimator for high-power noises or outliers. Due to such property, the ASE estimator is more efficient for trimming outliers that has long-tailed distributions \cite{andrews1974robust,black1996unification,maronna2019robust}. The first new algorithm is termed IWF-ASE, which combines the IWF with the ASE. Although the IWF-ASE algorithm achieves improved performance, the computational load of the algorithm can be further reduced. By applying the DCD method to solve the batch solution, the DCD-ASE algorithm is developed, which results in performance comparable to that of the IWF-ASE algorithm and has lower computational cost. As an added contribution, we introduce the proposed algorithms for the partial discharge (PD) denoising problem. Experimental studies are carried out to demonstrate the effectiveness of the proposed algorithms.

%\vspace{-5mm}
\section{Preliminaries}
\label{sec:format}
Considering a system identification problem, where the adaptive filter is utilized to model an unknown system. The desired signal at time instant $n$ can be expressed by the linear regression model
\begin{equation}
d(n) = {\bm w}_o^{\mathrm T}{\bm x}(n) + \xi(n)
\label{001}
\end{equation}
where $\bm w_o$ denotes an unknown impulse response of systems (of length $L$) to be estimated, $\bm x(n)=[x(n),x(n-1),\ldots,x(n-L+1)]^{\mathrm T}$ denotes the input vector, and $\xi(n)$ denotes the system noise. The filter output is $y(n) = {\bm w}^{\mathrm T}(n-1){\bm x}(n)$, where ${\bm w}(n-1)$ is the weight vector at time instant $n-1$. Therefore, the error signal is defined by $e(n) \triangleq d(n) - y(n)$ to update $\bm w(n)$ in a recursive manner.

\section{Proposed algorithms}
\subsection{IWF-ASE algorithm}
\begin{figure}[!htb]
	\centering
	\includegraphics[scale=0.34] {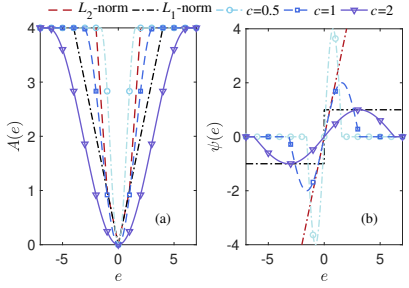}
	\caption{\label{1} Cost and score functions of the $L_2$-norm, $L_1$-norm, and ASE.}
	\label{Fig01}
\end{figure}

The robust ASE is defined as
\begin{equation}
{A}(e) \triangleq \left\{ \begin{array}{l}
4\sin^2\left(\frac{e}{2c}\right)\;\;\;{\rm {if}}\;|e| \leq \pi c \\ 
4\;\;\;\;\;\;\;\;\;\;\;\;\;\;\;\;\;\;{\rm {otherwise}} \\ 
\end{array} \right.
\label{004}
\end{equation}
where $c$ is a positive constant that controls the shape of the loss function. Fig. \ref{Fig01} shows the score functions $\psi(e)$ of the ASE, where 
\begin{equation}
\begin{aligned}
\psi(e)= \partial{A}(e)/\partial e 
= \left\{ \begin{array}{l}
\frac{2}{c}\sin\left(\frac{e}{c}\right).\;\;\;{\rm {if}}\;|e| \leq \pi c \\ 
0\;\;\;\;\;\;\;\;\;\;\;\;\;\;\;\;{\rm {otherwise}} \\ 
\end{array} \right.
\label{005}
\end{aligned}
\end{equation}
One can observe that ${A}(e)$ is a nonnegative scalar. The $L_1$-norm has a higher variance than the $L_2$-norm for normally distributed data. Hence, the ASE is designed to be similar to quadratic ($L_2$-norm) for small $e$ to achieve fast convergence \cite{andrews1974robust}. The ASE behaves similar to the $L_1$-norm for large $e$ and thus the algorithm achieves stable performance in the presence of various outliers. For very large $e$, the ASE becomes zero. To accelerate the convergence of the LMS algorithm, it is natural to exploit the least square (LS) criterion. Under this criterion, the cost function of the IWF-ASE algorithm is defined by 
\begin{equation}
{\mathcal J}(n) \triangleq \sum\limits_{i=1}^n {\lambda^{n-i}{A}(e(i,n))}  
\label{006}
\end{equation}
where $0\ll\lambda<1$ is the forgetting factor and the error signal $e(i,n)$ is defined as $e(i,n) \triangleq d(i) - \bm x^{\mathrm T}(i)\bm w(n)$ \cite{navia2011combination}. Taking the gradient of ${\mathcal J}(n)$ with respect to the weight vector $\bm w(n)$, and letting the equation be zero, we obtain
\begin{equation}
\sum\limits_{i=1}^n \lambda^{n-i}\phi(i,n)\bm x(i){\bm x^{\mathrm T}}(i)\bm w(n) = \sum\limits_{i=1}^n {\lambda^{n-i}\phi(i,n)\bm x(i)d(i)}
\label{008}
\end{equation}
where 
\begin{equation}
\phi(i,n) = \left\{ \begin{array}{l}
\frac{2}{c}\frac{\sin\left(\frac{e(i,n)}{c}\right)}{e(i,n) + \zeta}\;\;\;\;\;\;\;{\rm {if}}\;|e(i,n)| \leq \pi c \\ 
0\;\;\;\;\;\;\;\;\;\;\;\;\;\;\;\;\;\;\;\;\;\;{\rm {otherwise}} \\ 
\end{array} \right.
\label{009}
\end{equation}
is the weighting factor and $\zeta=0.0001$ is the regularization parameter. The expression of $\bm w(n)$ can be rewritten as $\bm w(n) = {\bm R}^{-1}(n){\bm \theta}(n)$, where $\bm R(n) = \sum_{i = 1}^n \lambda^{n-i}\phi(i,n)\bm x(i){\bm x^{\mathrm T}}(i)$
and $\bm \theta(n) = \sum_{i = 1}^n \lambda^{n-i}\phi(i,n)d(i)\bm x(i)$. If $\phi(i,n) = 1$, the above update equation becomes the fixed-point algorithm. If $\phi(i,n)\ne1$, $\bm R(n)$ and ${\bm \theta}(n)$ are the weighted autocorrelation matrix and the weighted cross-correlation vector of the optimal weights through $\phi(i,n)$. We have to recalculate $\bm w(n)$ at each iteration. In our previous studies, an online recursive method is considered to overcome this limitation \cite{lu2018recursive}. By using this approach, $\bm R(n)$ and $\bm \theta(n)$ can be approximated by the following equations:
\begin{equation}
\begin{aligned}
\bm R(n) \approx& \sum\limits_{i = 1}^n \lambda^{n-i}\phi(i,i)\bm x(i){\bm x^{\mathrm T}}(i) \\
=&\; \lambda{\bm R}(n-1) + \phi(n,n)\bm x(n){\bm x^{\mathrm T}}(n),\\
\end{aligned}
\label{011}
\end{equation}
and
\begin{equation}
\begin{aligned}
{\bm \theta}(n) \approx& \sum\limits_{i = 1}^n \lambda^{n-i}\phi(i,i)d(i)\bm x(i) \\
=&\; \lambda {\bm \theta}(n-1) + \phi(n,n)\bm x(n)d(n).
\end{aligned}
\label{012}
\end{equation}
\noindent\textit{Remark 1}: It is worth noticing that the IWF-ASE algorithm can further reduce the computational complexity. If $|e(i,n)|$ is larger than $\pi c$, $\phi(i,n)=0$, i.e., (\ref{011})-(\ref{012}) will not be updated. The \textit{update ratio} (UR) of the IWF-ASE algorithm can be different for various scenarios.

Because $\bm R(n)$ is self-adjoint, the solution can be regarded as the minimizer of the following quadratic form
\begin{equation}
\begin{aligned}
{\mathrm {E}}\{\bm w(n-1)\} =&\; \frac{1}{2}\bm w^{{\mathrm {T}}}(n-1)\bm R(n-1)\bm w(n-1) \\
&- \bm w^{{\mathrm {T}}}(n-1)\bm \theta(n-1).
\end{aligned}
\label{013}
\end{equation}
Thus, the gradient of the above equation is $\nabla{\mathrm {E}}\{\bm w(n-1)\} = \bm \theta(n-1) - \bm R(n-1)\bm w(n-1)
=\bm r(n-1)$. In this expression, the vector $\bm r(n)$ can also be interpreted as the residual. Following the iterative procedure, it is necessary to design $\bm w(n)$ so that it converges to $\bm w_o$. Based on the steepest descent method, the weight adaptation can be given by 
\begin{equation}
\begin{aligned}
\bm w(n) =&\; \bm w(n-1) + \mu(n-1) \left[\bm \theta(n-1) - \bm R(n-1)\bm w(n-1) \right]\\
=&\; \bm w(n-1) + \mu(n-1)\bm r(n-1)
\end{aligned}
\label{015}
\end{equation}
where $\mu(n-1)$ is the VSS of the IWF-ASE algorithm. Minimizing (\ref{013}), the VSS $\mu(n)$ can be obtained by
\begin{equation}
\begin{aligned}
\mu(n-1) = \frac{\bm r^{{\mathrm {T}}}(n-1)\bm r(n-1)}{\bm r^{{\mathrm {T}}}(n-1)\bm R(n-1)\bm r(n-1)}.
\end{aligned}
\label{016}
\end{equation}

\subsection{DCD-ASE algorithm}

Considering the initial regularization of the autocorrelation matrix $\bm R(n)$, i.e., $\bm R(0)=\rho {\bf I}$, where $\rho>0$ denotes the positive constant and ${\bf I}$ denotes the identity matrix. In this setting, (\ref{006}) can be rewritten as the following problem:
\begin{equation}
\bm w(n) = {\rm {arg}}\min_{\bm w} \left\{\sum\limits_{i=1}^n {\lambda^{n-i}{A}(e(i,n))} + \delta(n) \left\| \bm w\right\|_2^2\right\}
\label{017}
\end{equation}
where $\left\| \cdot \right\|_2$ denotes the $L_2$-norm, and $\delta(n)=\lambda^{n+1}\rho$ denotes the time-varying regularization factor, which decreases exponentially over time. Reusing expression of  $\bm w(n)$, $\bm R(n)$ can be rewritten as
\begin{equation}
\begin{aligned}
\bm R(n) =&\;\lambda{\bm R}(n-1) + \phi(n,n)\bm x(n){\bm x^{\mathrm T}}(n) \\
&+ \left(\delta(n)-\lambda\delta(n-1)\right){\bf I}.
\end{aligned}
\label{018}
\end{equation}

Defining $\hat {\bm w}(n-1)$ being the estimate of $\bm w(n-1)$ at time instant $n-1$, and the residual vector can be rewritten as $\bm r(n-1)=\bm \theta(n-1) - \bm R(n-1)\hat {\bm w}(n-1)$. Then, defining the incremental vector of the weight vector $\Delta\bm w(n)=\bm w(n)-\hat{\bm w}(n-1)$, the auxiliary equation $\bm R(n)\Delta \bm w(n) = \bm\theta(n)-\bm R(n)\hat{\bm w}(n-1)$ can be obtained. Combining (\ref{012}) and (\ref{018}), we obtain
\begin{equation}
\begin{aligned}
\bm \zeta(n) =&\; \lambda\bm r(n-1)+\phi(n,n)e(n)\bm x(n) \\
&- \left(\delta(n)-\lambda\delta(n-1)\right) \hat{\bm w}(n-1)
\end{aligned}
\label{020}
\end{equation}
By using the leading DCD method, we can minimize (\ref{017}) without resorting to the matrix inversion, which is formulated as
\begin{equation}
\begin{aligned}
\hat{\bm w}(n)=\hat{\bm w}(n-1)+\Delta\hat{\bm w}(n).
\end{aligned}
\label{021}
\end{equation}
Here, $\Delta\hat{\bm w}(n)$ is obtained by the leading DCD method, as presented in Table \ref{Table01}, where $[\bm r(n)]_l$ denotes the $l$th entry of $\bm r(n)$, $[\bm R(n)]_{l,l}$ represents the $l$-$l$th element, and $[\bm R(n)]_{:,l}$ stands for the $l$th column of $\bm R(n)$\footnote{For more details, readers can refer to \cite{zakharov2008low,yu2019dcd,2008Low}.}. Herein, $[-H,H]$ denotes the assumed range of the solution vector and the step size $m$ is quantized by $H\frac{1}{2}, H\frac{1}{2^2}, \ldots, H\frac{1}{2^{M_b}}$, where $M_b$ denotes the number of bits for a fixed-point representation of $\hat{\bm w}(n)$ within the range of $[-H,H]$, and $N_u$ denotes the maximum number of iteration. The accuracy of $\hat{\bm w}(n)$ is increased as $N_u$ increases. At each iteration, $\hat{\bm w}(n)$ approaches the optimal one with shift and addition operations. As such, multiplications and divisions in RLS and IWF-ASE can be implicitly avoided. 

\begin{table}[tbp]
	\scriptsize
	\centering
	\caption{Leading DCD method.}
	\doublerulesep=0.5pt
	\begin{tabular}{lll}
		\cline{1-1}
		\multicolumn{1}{l}{\begin{tabular}[c]{@{}l@{}}\textbf{Input}: $\bm R(n)$,\;$\bm \zeta(n)$,\;$L$,\;$H$,\;$N_u$, \textbf{and} $M_b$ \\ \textbf{Initialization}: $\Delta\hat{\bm w}(n)\gets\bm 0$, $\bm r(n)\gets\bm \zeta(n)$, $q \gets 1$, \textbf{and} $m \gets \frac{H}{2}$ \end{tabular} } &  &  \\ \cline{1-1}
		\multicolumn{1}{l}{\begin{tabular}[c]{@{}l@{}} \texttt{For $j=1,2,\ldots,N_u$} \\
				\;\;\;$l={\rm {arg}}\max\limits_{j=1,2,\ldots,L}\left\{ |[\bm r(n)]_j|\right\}$\\
				\;\;\;\texttt{While $|[\bm r(n)]_l| \leq \frac{m}{2}[\bm R(n)]_{l,l}$\;\;and\;$q \leq M_b$}\\
				\;\;\;\;\;\;$q \gets q+1$,\;$m \gets \frac{m}{2}$;\\
				\;\;\;\texttt{End While}\\
				\;\;\;\texttt{If $q > M_b$}\\
				\;\;\;\;\;\;\texttt{Break};\\
				\;\;\;\texttt{Elseif}\\
				\;\;\;\;\;\;$[\Delta\hat {\bm w}(n)]_l \gets [\Delta\hat {\bm w}(n)]_l + m{\rm {sign}}([\bm r(n)]_l)$; \\
				\;\;\;\;\;\;$\bm r(n) \gets \bm r(n) - m{\rm {sign}}([\bm r(n)]_l)[\bm R(n)]_{:,l}$; \\
				\;\;\;\texttt{End If}\\
		\texttt{End For}\end{tabular}} &  &  \\ \cline{1-1}
		&  & 
	\end{tabular}
\label{Table01}
\end{table}

\noindent\textit{Remark 2}: The DCD-ASE algorithm still retains the select updated property as we illustrated in \textit{Remark 1}. Such merits can promote the algorithm to further reduce the complexity. Therefore, the DCD-ASE algorithm has lower computational complexity as compared to the algorithm in \cite{zhang2016convex,yu2019dcd}. 

\begin{table}[tbp]
	\scriptsize
	\centering
	\caption{Summary of the computational complexities of the algorithms in each iteration.}
	\doublerulesep=0.5pt
	\begin{tabular}{l|l|l|l}
		\hline
		\textbf{Algorithm}                                                    & \multicolumn{1}{c|}{\textbf{$+/-$}} & \multicolumn{1}{c|}{\textbf{$\times / \div$}} & \textbf{Other operations}\\ \hline
		\textbf{RMCC}\cite{zhang2016convex}                                                         &$3L^2+7L+4$     &$3L^2+13L+12$    &\multicolumn{1}{c}{\begin{tabular}[c]{@{}l@{}}Exponential \\ operation\end{tabular}}            \\ \hline
		\textbf{IWF}\cite{xi2013iterative}                                                          &$3L^2+4L$     &$3.5L^2+6L$      &\multicolumn{1}{c}{--}          \\ \hline
		\textbf{DCD-RMCC}\cite{yu2019dcd}                    &\multicolumn{1}{c|}{\begin{tabular}[c]{@{}l@{}}$3L+2N_uL$\\$+M_b$ \end{tabular}}     &$7L+5$     &\multicolumn{1}{c}{\begin{tabular}[c]{@{}l@{}}Exponential \\ operation \end{tabular}}        \\ \hline
		\textbf{IWF-ASE}&$3L^2+4L$     &$3.5L^2+8L+3$       &\multicolumn{1}{c}{\begin{tabular}[c]{@{}l@{}}Comparison and \\ sine operation\end{tabular}}          \\ \hline
		\textbf{DCD-ASE}&\multicolumn{1}{c|}{\begin{tabular}[c]{@{}l@{}}$3L+2N_uL$\\$+M_b+1$ \end{tabular}}    &$7L+6$       &\multicolumn{1}{c}{\begin{tabular}[c]{@{}l@{}}Comparison and \\ sine operation\end{tabular}}          \\ \hline
	\end{tabular}
\label{Table02}
\end{table}

Table \ref{Table02} lists the computational complexities of existing algorithms and the proposed algorithm. From (\ref{009}), (\ref{011}), (\ref{012}), and (\ref{015}), it is clear that the IWF-ASE algorithm requires additional $2L+3$ multiplications, 1 comparison, and sine calculation when compared to the conventional IWF algorithm. It still has an affordable computational load for implementation.  It is generally known that the conventional DCD-RLS algorithm requires $3L+2N_uL+M_b$ additions and $5L+2$ multiplications \cite{yu2019dcd}. The DCD-RMCC algorithm requires $3L+2N_uL+M_b$ additions, $7L+5$ multiplications, and exponential operation. The DCD-ASE algorithm requires additional 1 addition and $2L+4$ multiplications for computing ASE. It requires $3L+2N_uL+M_b+1$ additions and $7L+6$ multiplications, which significantly reduces the computational complexity as compared to the IWF-ASE algorithm.

\section{Illustrative Examples}
In this section, two examples are presented to demonstrate the improved performance of the proposed algorithms. The impulsive noise is modeled by Bernoulli-Gaussian (BG) distribution \cite{bhotto2012new,papoulis2004normalized,liang2020performance}. The normalized mean
square deviation (NMSD) ${\rm {NMSD}}(n)=10\log\{ \left\| \bm w(n)-\bm w_o\right\|_2^2 / \left\| \bm w_o\right\|_2^2 \}$ and mean square error (MSE) are used as evaluation. All the results are averaged over 100 Monte-Carlo runs.
\subsection{Example A: System identification}
\begin{figure}[!htb]
	\centering
	\includegraphics[scale=0.34] {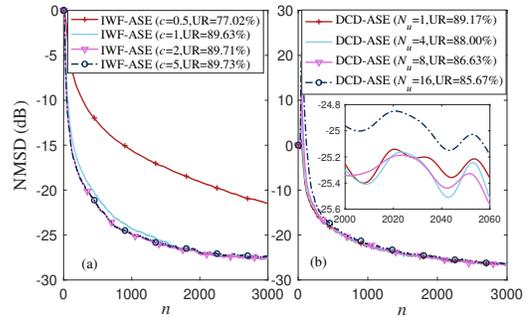}
	\caption{\label{2} The proposed algorithms versus different parameter settings in impulsive scenario ($\lambda=0.999$, $\rho=0.0001$).}
	\label{Fig02}
\end{figure}
\begin{figure}[!htb]
	\centering
	\includegraphics[scale=0.38] {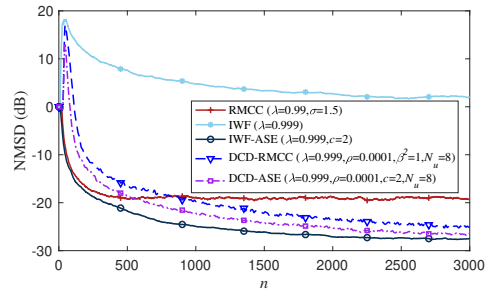}
	\caption{\label{3} Comparison of the algorithms in impulsive scenarios.}
	\label{Fig03}
\end{figure}
In this example, the unknown system is a ten-tap system, which is generated randomly. The input signal is a white Gaussian noise (WGN) with zero mean and unit variance. In impulsive scenarios, the probability of occurrence for impulsive samples is $P_r=0.1$, the variance is $\sigma_\xi^2={10^4}$, and the WGN with signal-to-noise ratio (SNR)$=0$dB is used.

First, the effect of the parameter settings on the proposed algorithms is investigated in Fig. \ref{Fig02}. One can observe that the IWF-ASE algorithm can obtain good performance for $c=1,2$, and 5. Considering the UR and performance, $c=2$ is set to the DCD-ASE algorithm. Then, the NMSDs of DCD-ASE algorithm with different $N_u$ are shown in Fig. \ref{Fig02} (b), where $H=2$ and $M_b=8$. It can be observed that all the parameter settings have similar convergence performance and UR. In the following simulations, $N_u=8$ is selected.

Next, a comparison of the algorithms is shown in Fig. \ref{Fig03}, where the RMCC \cite{zhang2016convex}, IWF \cite{xi2013iterative},  and DCD-RMCC \cite{yu2019dcd} algorithms are used as the benchmark. With a similar convergence rate, the proposed algorithms can achieve smaller misalignment in impulsive noise. At the price of slightly slow convergence, the low-cost DCD-ASE algorithm has similar performance to the IWF-ASE algorithm.

\subsection{Example B: Partial discharge denoising application}
\begin{figure}[!htb]
	\centering
	\includegraphics[scale=0.6] {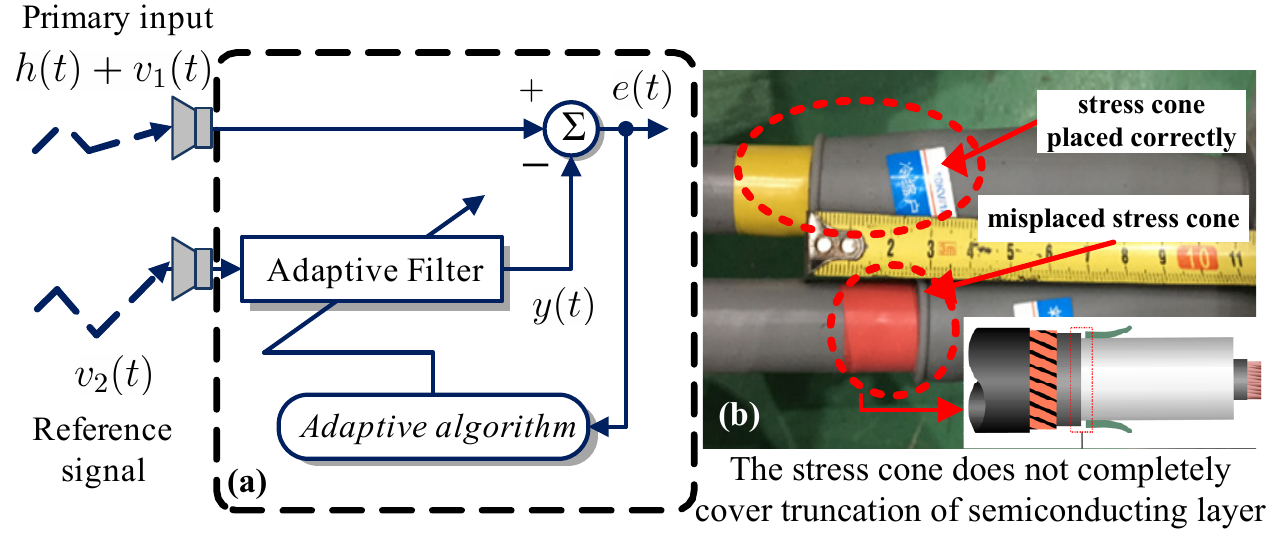}
	\caption{\label{4} (a) Adaptive noise cancellation system. (b) Photo of the XLPE cable termination with the defect of mismatch in prefabricated part.}
	\label{Fig04}
\end{figure}
\begin{figure}[!htb]
	\centering
	\includegraphics[scale=0.4] {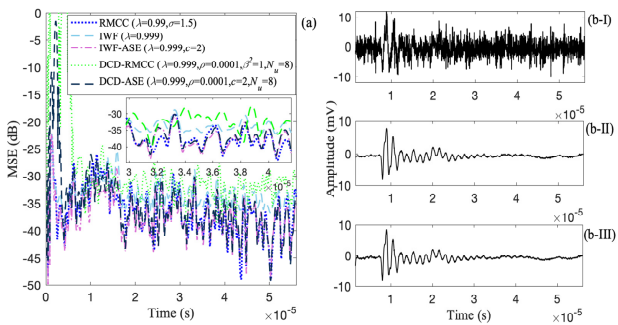}
	\caption{\label{5} (a) MSEs of the algorithms in impulsive scenarios. (b) Experimental results for the proposed algorithms in impulsive scenarios: (b-I) Noisy PD signal; (b-II) filtered PD signal by IWF-ASE; (b-III) filtered PD signal by DCD-ASE.}
	\label{Fig05}
\end{figure}

In the second example, the proposed algorithms are utilized to the adaptive noise cancellation system for PD denoising. Fig. \ref{Fig01} (a) shows the configuration of adaptive noise cancellation system, where $h(t)$ represents the pure PD signal at time $t$, $v_1(t)$ denotes the noise signal, and $v_2(t)$ stands for the reference signal of adaptive filter. The input vector $\bm x(t)$ for impulsive scenario is generated by the impulsive signal through a low-pass infinite impulse response (IIR) filter, where the impulsive signal is generated by a signal generator, which can be modeled by BG distribution with $Pr = 0.1$ and $\sigma_\xi^2 = 25$. The transfer function of IIR filter is ${\mathcal N}^{-1}(z)=1-0.2z^{-1}$ \cite{gorriz2009a}. Before that, some adaptive filtering algorithms have been developed for PD denoising  \cite{hariri1996field,mohammadirad2018localization,lu2019affine,lu2021tde}. We repeat the same experiments as \cite{lu2021tde}, but this time using the proposed algorithms in impulsive scenarios and the length of the weight vector is set to 5. The cross-linked polyethylene (XLPE) cable termination is selected as an object to verify the effectiveness of the algorithm. A corona-free testing transformer is exploited to the test platform. The capacitive divider offers a circuit loop for the PD current which provides a power-frequency voltage reference phase. The PD current is transmitted to the oscilloscope (Rigol DS6104) via the coaxial cable by employing the high-frequency current transformers (HFCT). The bandwidth of HFCT is 2.5-216MHz. The maximum voltage of the transformer without PD is 150kV and its capacity is 75kVA. The PD inception voltage is 8.7kV and the test voltage is set to 17kV. The voltage is increased at a boosted
rate of 1kV/s until a PD occurs. During the manufacturing of the cable termination, the local electric field at the cutoff point of the semiconductive layer will be distorted, which leads to the discharge phenomena. To simulate the defect of mismatch in the prefabricated part, the stress cone is fixed 30mm backward (see Fig. \ref{Fig04} (b)).

Fig. \ref{Fig05} (a) plots the MSEs of the algorithms. All the parameter settings are the same as Example A. We can see that the IWF algorithm has a large MSE in impulsive noise. In contrast, the IWF-ASE and DCD-ASE algorithms have improved performance. With slightly slow convergence, the DCD-ASE algorithm significantly reduces the computational complexity and maintains the steady-state performance. To further demonstrate the effectiveness of the proposed algorithms, Fig. \ref{Fig05} (b) shows the experimental results of the algorithms. It can be observed from this figure that the IWF-ASE and DCD-ASE algorithms can effectively remove noise. Moreover, one can see that the PD signal after IWF-ASE and DCD-ASE filtering
completely meets the IEEE Std 400.3-2006 \cite{standard2000high}, i.e., the noise
level is less than a specified permissible PD magnitude.
%\vspace{-5mm}
\section{Conclusion}

In this paper, two algorithms that employ the ASE have been proposed. The classical IWF is difficult to converge in impulsive noise. In contrast, the IWF-ASE algorithm can attenuate the effect of the system noises and as such, it can achieve improved performance in various scenarios. To further reduce the computational burden, the DCD-ASE algorithm has been developed. By making use of the DCD method, the DCD-ASE algorithm significantly reduces the computational complexity and maintains the performance. Simulations and experiments performed in the context of system identification and PD denoising have been proved the validity of the proposed algorithms in impulsive noise.

\ifCLASSOPTIONcaptionsoff
\newpage
\fi

\footnotesize
\bibliographystyle{IEEEtran}
\bibliography{IEEEabrv,mybibfile2}

\end{document}